\documentclass[letterpaper,english,aps,pre, preprint]{revtex4}
\usepackage[T1]{fontenc}
\usepackage[latin1]{inputenc}
\usepackage{graphicx}

\makeatletter

\renewcommand{\@dotsep}{100}
\renewcommand{\@pnumwidth}{1000mm}

\usepackage{babel}
\makeatother
\begin{document}

\title{Resonance Clustering in Globally Coupled Electrochemical Oscillators
with External Forcing}

\author{István Z. Kiss}

\affiliation{Department of Chemistry, 3501 Laclade Ave., Saint Louis University,
St. Louis, MO 63103}

\author{Yumei Zhai, and John L. Hudson}

\email{hudson@virginia.edu}

\affiliation{Department of Chemical Engineering, 102 Engineers' Way, University
of Virginia, Charlottesville, VA 22904-4741}

\begin{abstract}
Experiments are carried out with a globally coupled, externally forced
population of limit-cycle electrochemical oscillators with an approximately
unimodal distribution of heterogeneities. Global coupling induces
mutually entrained (at frequency $\omega_{1}$) states; periodic forcing
produces forced-entrained ($\omega_{\mathrm{F}}$) states. As a result
of the interaction of mutual and forced entrainment, resonant cluster
states occur with equal spacing of frequencies that have discretized
frequencies following a resonance rule $\omega_{n}\cong n\omega_{1}-(n-1)\omega_{\mathrm{F}}$.
Resonance clustering requires an optimal, intermediate global coupling
strength; at weak coupling the clusters have smaller sizes and do
not strictly follow the resonance rule, while at strong coupling the
population behaves similar to a single, giant oscillator. 
\end{abstract}
\maketitle

\section{Introduction}

Rhythms, often generated as synchronization of oscillator populations\cite{winfreebook},
can also be exposed to external global forcing or feedback. The resulting
dynamical behavior depends on the types of oscillators (smooth, relaxation,
chaotic), topology of interactions (local, global, or network), and
on the magnitude of forcing frequencies relative to the inherent frequencies
of the oscillators. A plethora of interesting behavior has been observed
with forcing in reaction-diffusion systems or locally coupled, identical
oscillators: some examples include labyrinthine standing waves \cite{Petrov.Ouyang.ea:97:Nature:Resonant-pattern-formation-in},
standing wave patterns\cite{Oertzen.Rotermund.ea:00:Journal-Of-Physical-Chemistry:Standing-wave-patterns-in},
resonant phase patterns\cite{Lin.Bertram.ea:00:Phys-Rev-Lett:Resonant-phase-patterns-in},
Bloch-front turbulence\cite{Marts.Hagberg.ea:04:Phys-Rev-Lett:Bloch-front-turbulence-in-periodically},
localized clusters\cite{Vanag.Zhabotinsky.ea:01:Phys-Rev-Lett:Oscillatory-clusters-in-periodically},
spirals with hypocycloidal trajectories\cite{Steinbock.Zykov.ea:93:Nature:Control-of-spiral-wave-dynamics},
wave traps and twisted spirals\cite{Rudzick.Mikhailov:06:Phys-Rev-Lett:Front-reversals-wave-traps},
and pacemaker entrainments\cite{Fukuda.Tamari.ea:05:J-Phys-Chem:Entrainment-in-chemical-oscillator}. 

Less attention has, however, been paid to the analysis of heterogeneous
populations of oscillators with a distribution of natural frequencies.
A simple example is a globally coupled population with global forcing.
Following Kuramoto's heuristic argument\cite{kuramotobook}, it is
expected that global coupling would produce a mutually entrained state;
this state behaves as a giant oscillator and becomes entrained to
the frequency of an external forcing signal. However, the frequency
adjustment process of the mutually entrained state is not trivial:
Sakaguchi \cite{Sakaguchi:88:Progress-Of-Theoretical-Physics:Cooperative-phenomena-in-coupled}
studied the effects of the external fields on mutual entrainment by
analysis of a phase model with unimodal heterogeneities. The simulation
results show that a transition from the mutual entrainment to the
forced entrainment occurs as forcing strength is enhanced; in between,
two and more plateaus are seen in the frequency-vs-natural-frequency
plots which indicate the formation of multiple, resonance clusters.
Because of the interaction of the mutually entrained cluster (with
frequency of $\omega_{1}$) and the forced cluster ($\omega_{\mathrm{F}}$),
new resonant clusters with frequency of $\omega_{n}\cong n\omega_{1}-(n-1)\omega_{f}$
are formed at weak forcing strengths. Note that these discretized
frequencies are equally spaced with $\Delta\omega=\omega_{1}-\omega_{f}$.
It was also shown that the order parameter exhibits large amplitude
oscillations when two major clusters are formed. \cite{Sakaguchi:88:Progress-Of-Theoretical-Physics:Cooperative-phenomena-in-coupled,Montbrio:Phys.-Rev.:2004:Synchronization} 

In this paper we investigate experimentally resonance clustering in
a chemical system (the oscillatory electrodissolution of a nickel
electrode array in sulfuric acid solution) with global coupling and
forcing with limit cycle oscillators. The mutual and forced entrainment
states are identified, and their interactions are analyzed. Features
of resonance clustering are compared as coupling strength and forcing
amplitude and frequency are varied. Numerical studies are carried
out to confirm the experimental finding of resonance clustering in
ordinary differential equation models and to investigate the features of the dynamics
in a large parameter space.

\section{Experiments}

\subsection{Experimental Setup}

The experiments were carried out in a standard three electrode electrochemical
cell containing 3 mol/dm$^{3}$ sulfuric acid at 11 $\mathrm{^{o}C}$
with Ni working, a $\mathrm{Hg}/\mathrm{Hg}_{2}\mathrm{SO}_{4}/\mathrm{K}_{2}\mathrm{SO}_{4}$
reference, and a Pt counter electrodes \cite{Kiss.Wang.ea:99:J-Phys-Chem-B:Experiments-on-arrays-of,Kiss.Zhai.ea:02:Science:Emerging-coherence-in-population}.
The currents ($i_{k}(t)$) of Ni electrodes (64 1-mm diameter electrodes
in an $8\times8$ geometry with 2 mm spacing) were measured at 100
Hz. The potential of each electrode was held at potential $V$ versus
$\mathrm{Hg/Hg_{2}\mathrm{SO_{4}}}/\mathrm{concentrated}\,\mathrm{K}_{2}\mathrm{SO_{4}}$
reference electrode. The external forcing is added to the applied
potential $V(t)=V_{0}+b\mathrm{sin}(2\pi f_{\mathrm{F}}t)$. $V_{0}=1.09$
V. The electrodes were connected to the potentiostat through one series
(collective) resistor, $R_{\mathrm{s}}$, and through 64 parallel
resistors\cite{Kiss.Wang.ea:99:J-Phys-Chem-B:Experiments-on-arrays-of,Kiss.Zhai.ea:02:Science:Emerging-coherence-in-population}.
The interaction strength $K$ was controlled through the external
resistors $K=(R_{\mathrm{s}}/R_{\mathrm{tot}})/(1-R_{\mathrm{s}}/R_{\mathrm{tot}})$
where $R_{\mathrm{tot}}=10.1\Omega$ was the total resistance.

The phases and frequencies of the oscillators were determined with
the Hilbert transform method \cite{sync_book} from time series data
of current $i_{k}(t)$ \cite{Kiss.Zhai.ea:02:Science:Emerging-coherence-in-population}.
An order parameter, defined as the normalized vector sum of the phase
points ($P_{k}(t)$) in $H(i_{k}(t)-<i>)$ vs. ($i_{k}(t)-<i>$) space,
is used to characterize the extent of the synchrony of the population\begin{equation}
Z(t)=\frac{\sum P_{k}(t)}{\sum_{k}|P_{k}(t)|},\label{eq: order_Z_kuramoto_calc}\end{equation}
where $H$ is the Hilbert transform. This order parameter \cite{Kiss.Zhai.ea:02:Science:Emerging-coherence-in-population,Zhai.Kiss.ea:04:Industrial-Engineering-Chemistry-Research:Emerging-coherence-of-oscillating,Zhai.Kiss.ea:04:Extracting-order-parameters-from-global}
is similar to the Kuramoto order parameter \cite{kuramotobook}. The
magnitude of $r=|Z|$, the order, has a maximum value of 1 for full
synchronization and zero for complete desynchronization (for a population
of infinite size).

\subsection{Results}

The population in this study has a nearly unimodal natural frequency
distribution (without coupling and forcing) with a standard deviation
of 8.5 mHz; %
\begin{figure}
\includegraphics[%
  width=0.9\columnwidth]{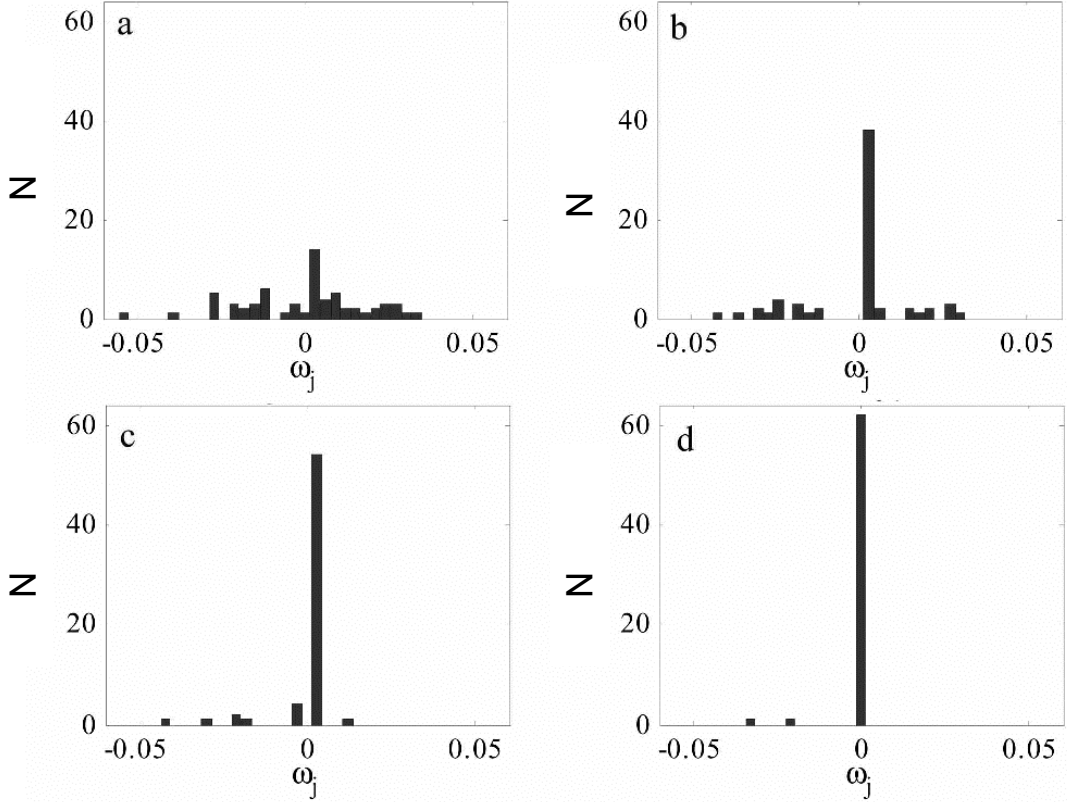}

\caption{\label{fig:exp_hist}Experiments: Dimensionless frequency histograms
($\omega=f/<f>-1$) at different coupling strengths for a unimodal
population. The mean natural frequency $<f>=0.47$Hz, standard deviation
$\sigma=8.5$mHz. (a) $K=0$. (b) $K=0.026$. (c) $K=0.031$. (d)
$K=0.042$.}
\end{figure}
 the frequency histogram is shown in Fig.~\ref{fig:exp_hist}a. Figs.~\ref{fig:exp_hist}b-d
also show the frequency histograms at $K=0.026,$ 0.031 and 0.042,
respectively. At $K=0.026$ (Fig.~\ref{fig:exp_hist}b), a dominant,
mutually entrained cluster emerged at approximately the mean natural
frequency. Elements with high and low natural frequencies were not
entrained yet. With a stronger coupling of $K=0.031$ (Fig.~\ref{fig:exp_hist}c),
the cluster grew in size considerably and only a few elements were
left out. At $K=0.042$ (Fig.~\ref{fig:exp_hist}d), 62 out of the
64 elements had been in the same cluster with two elements of lower
natural frequencies desynchronized. External periodic forcing was
applied to all these three partially synchronized states. 

First consider the forcing of the most synchronized state at $K=0.042$.
\begin{figure}
\includegraphics[%
  width=0.9\columnwidth]{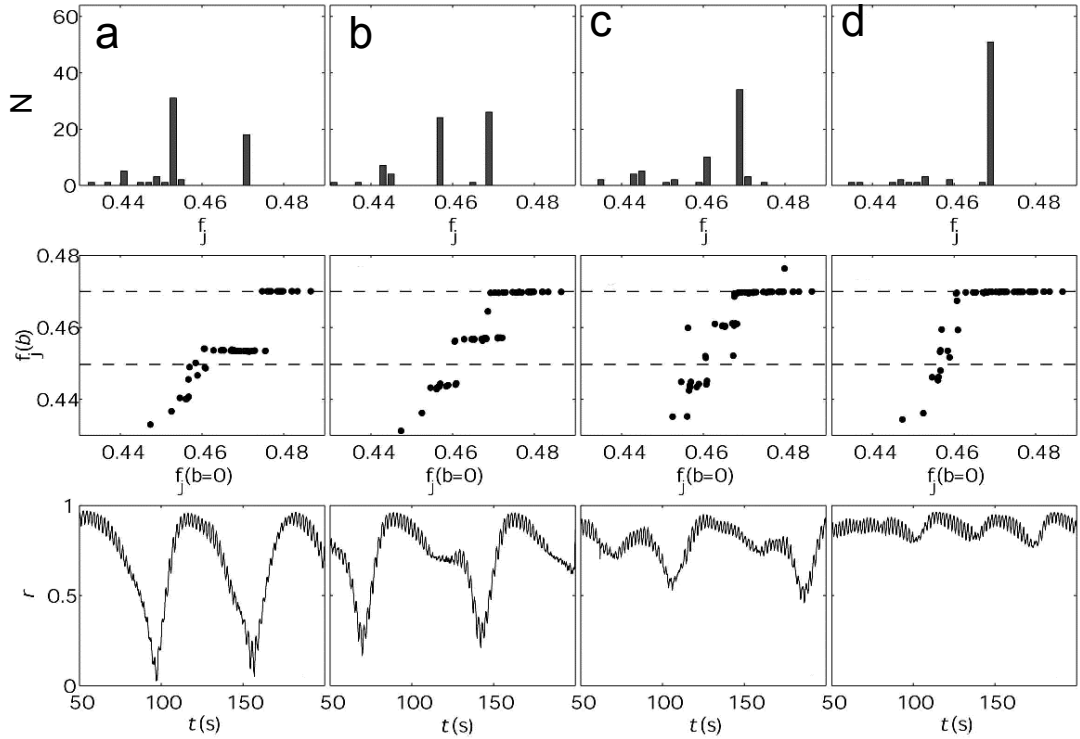}

\caption{\label{fig:exp_resonance}Experiments: Periodic forcing on a unimodal
population with $K=0.042$. $V_{0}=1.09$V. Mean frequency before
forcing $<f>=0.4497$ Hz, forcing frequency $f_{\mathrm{F}}=0.47$Hz,
$f_{\mathrm{F}}-<f>>2\sigma$. Top row: Frequency histograms at different
forcing amplitudes. Middle row: Frequencies in presence of coupling
and forcing versus the corresponding natural frequencies. Upper and
lower dashed line are $f_{\mathrm{F}}$ and $<f>$, respectively.
Bottom row: Time series of order $r$. a. $b=3.3$mV. b. $b=4.0$mV.
c. $b=4.6$mV. d. $b=5.3$mV. }
\end{figure}
Without forcing a mutually entrained cluster of 62 elements formed
at $\omega_{1}=$0.450 Hz. The periodic forcing was applied at a higher
frequency of $\omega_{\mathrm{F}}=f_{\mathrm{F}}=0.47$Hz. With forcing
amplitude of $b=3.3$mV some elements of the original (mutually entrained)
cluster moved out and formed a new cluster at the forcing frequency
(forced entrainment). In Fig.~\ref{fig:exp_resonance}a (frequency
histogram, top panel) there were two large peaks corresponding to
the two clusters. The plot of frequencies-vs-natural-frequencies (middle
panel) also clearly shows two plateaus; the one of the forced entrainment
was at $f_{\mathrm{F}}$ while the other one of the mutual entrainment
had a frequency that is slightly higher than the mean frequency without
forcing. The order oscillated in large amplitudes at $b=3.3$mV (bottom
panel). As $b$ increased to 4.0mV, the forced entrained cluster grew
in size; simultaneously two mutual entrained clusters emerged (Fig.~\ref{fig:exp_resonance}b).
The frequency of the old mutual entrained cluster ($\omega_{\mathrm{l}}$)
further increased to be closer to $\omega_{\mathrm{F}}$. A new, small
resonant cluster formed at $\omega_{2}=2\omega_{1}-\omega_{\mathrm{F}}$.
With a even stronger forcing of $b=4.6$mV, more resonant clusters
formed at frequencies of $\omega_{n}\cong n\omega_{\mathrm{l}}-(n-1)\omega_{\mathrm{F}}$,
$n=2,3,4$ (Fig.~\ref{fig:exp_resonance}c). However, each of these
clusters had a smaller size compared with the two mutual entrained
clusters at $b=4.0$mV (Fig.~\ref{fig:exp_resonance}b). With only
one large cluster the order no longer exhibited large amplitude oscillations.
Note that the spacing between any adjacent clusters became smaller
as $b$ increased. When the forcing amplitude was further increased
to $b=5.3$mV, no distinct resonant clusters were observed and only
one large, the forced entrained cluster, appeared with 12 non-entrained
elements scattering in the lower frequency region (Fig.~\ref{fig:exp_resonance}d).
Finally, with strong enough forcing amplitude, all the elements formed
one cluster at the forcing frequency (not shown). 

\begin{figure}
\includegraphics[%
  width=0.9\columnwidth]{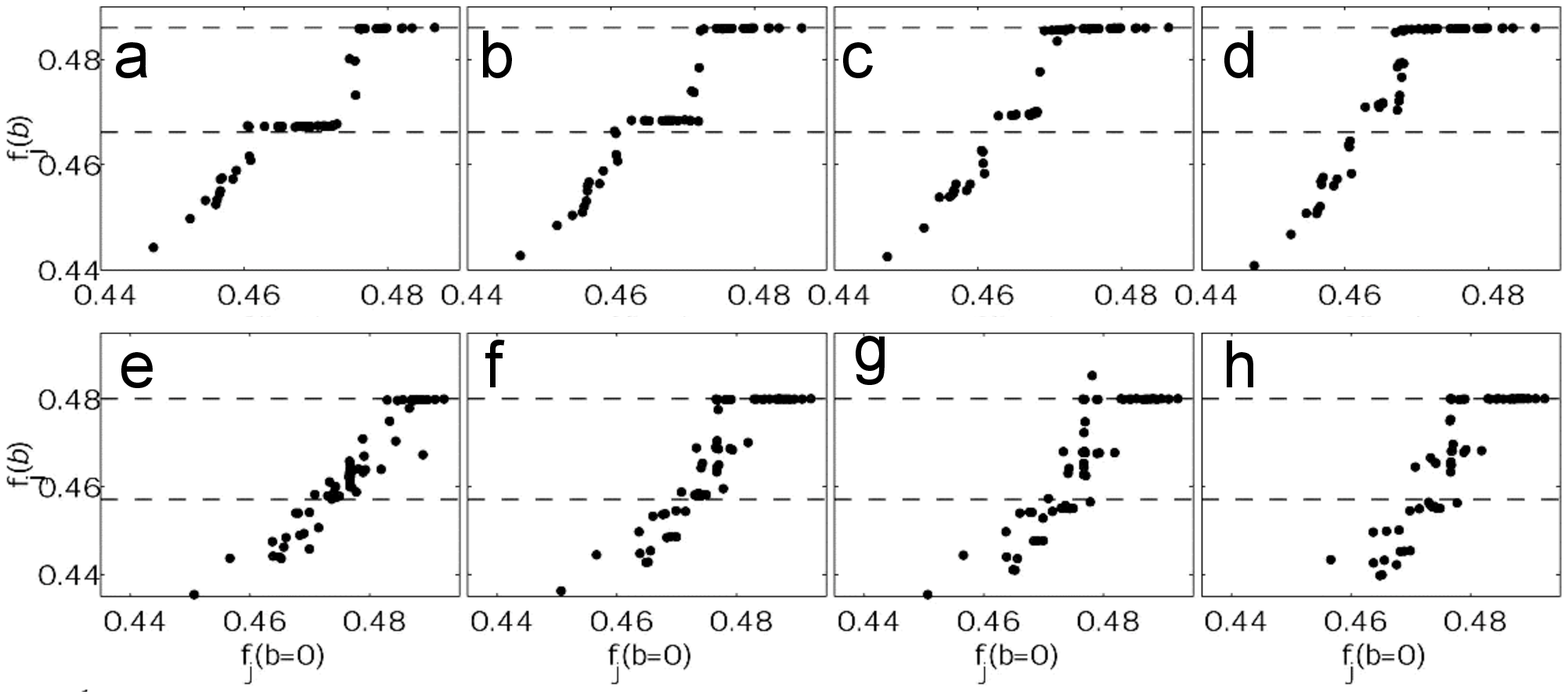}

\caption{\label{fig:exp_weak}Periodic forcing on a unimodal population with
weak coupling strengths. Frequencies in presence of coupling and forcing
versus the corresponding natural frequencies. Top row: $K=0.031$,
$<f>=0.466$Hz, $f_{\mathrm{F}}=0.486$Hz. a. $b=3.3$mV. b. $b=4.0$mV.
c. $b=4.6$mV. d. $b=5.3$mV. Bottom row: $K=0.026$, $<f>=0.4571$Hz,
$f_{\mathrm{F}}=0.48$Hz. e. $b=3.3$mV. f. $b=4.95$mV. g. $b=5.3$mV.
h. $b=5.95$mV. Upper and lower dashed line are $f_{\mathrm{F}}$
and $<f>$, respectively.}
\end{figure}

During the periodic forcing of a less synchronized state obtained
at $K=0.031$ (Fig.~\ref{fig:exp_hist}c), the occurrence of resonant
clusters requires stronger forcing strengths as shown in Figs.~\ref{fig:exp_weak}a-d.
 Forced entrainment occurred with forcing and two main clusters coexisted
at weak forcing strengths (Figs.~\ref{fig:exp_weak}a, b). As $b$
increases, the forced entrained cluster grew larger (Figs.~\ref{fig:exp_weak}a-d)
and finally became the only dominant cluster at $b>5.3$mV. The third,
new resonant cluster with the frequency of $\omega_{2}=2\omega_{\mathrm{l}}-\omega_{\mathrm{F}}$
appeared at $b=4.0$mV; however, compared with the case of $K=0.042$
at the same weak forcing strength (Fig.~\ref{fig:exp_resonance}f),
the new cluster was much smaller and can be barely seen (Fig.~\ref{fig:exp_weak}b).
The presence of the third cluster only became obvious at a greater
forcing amplitude of $b=4.6$mV (Fig.~\ref{fig:exp_weak}c); in contrast,
in the case of $K=0.042$ at $b=4.6$mV multiple resonant clusters
had emerged (Fig.~\ref{fig:exp_resonance}c). Because of the small
size of the resonant cluster, the existence of two main clusters at
$b=4.6$mV the order oscillated in large amplitudes (not shown). At
$b=5.3$mV, the forcing was strong enough to induce multiple resonant
clusters with closer equal frequency spacing (Fig.~\ref{fig:exp_weak}d)
and with smaller sizes. Again, before the system reached a fully forced
entrainment state (not shown in the figures) the mutual entrainment
clusters completely disappeared by giving some of their elements to
the forced entrainment cluster while leaving the others nonentrained. 

The forcing experiments were carried out for an even less synchronized
population of oscillators obtained with $K=0.026$ (Fig.~\ref{fig:exp_hist}b).
With a small forcing strength of $b=3.3$ mV (Fig.~\ref{fig:exp_weak}e),
a high peak arose at the forcing frequency in the frequency histogram
; however, the mutual entrained cluster had broken up because of the
weak coupling. In the frequency-vs-natural-frequency plot below the
forced entrained plateau, the frequencies of the nonentrained elements
formed an almost continuous $45^{\mathrm{o}}$ line in the lower frequency
region with a small shoulder around 0.4571 Hz (which is close to the
mean natural frequency of the oscillators). The order exhibited fairly
large oscillations but these oscillations were not regular (not shown).
As $b$ was increased to a quite strong value of 4.95mV, the elements
with lower frequencies started to form multiple small groups as shown
in Fig.~\ref{fig:exp_weak}f. However, these 'clusters' were small
and close to each other on the $45^{\mathrm{o}}$ line in the frequency-vs-natural-frequency
plot (Fig.~\ref{fig:exp_weak}f). Small clusters were also observed
at stronger forcing strengths $b=5.3$mV (Fig.~\ref{fig:exp_weak}g
) and $b=5.95$mV (Fig.~\ref{fig:exp_weak}h). Since there was no
second large cluster besides the forced entrained one, the order oscillations
were irregular with relatively small amplitudes (not shown). Finally,
a completely forced entrained cluster was formed with strong enough
forcing amplitudes.

\section{Numerical Simulations}

\subsection{Model Equations}

We used a model of anodic electrodissolution of a single nickel electrode
proposed by Haim et al. \cite{Haim.Lev.ea:92:Journal-Of-Physical-Chemistry:Modeling-periodic-and-chaotic}.
The model in a dimensionless form involves two variables: the dimensionless
double layer potential drop ($e$) and the surface coverage of NiO+NiOH
($\theta$). One oscillator is described by the following two equations:\begin{eqnarray}
\frac{de}{dt} & = & \frac{V-e}{R}-i_{\mathrm{F}}(\theta,e)\label{eq:ni_general}\\
\Gamma\frac{d\theta}{dt} & = & \frac{\mathrm{exp}(0.5e)}{1+C_{\mathrm{h}}\mathrm{exp}(e)}(1-\theta)-\frac{BC_{\mathrm{h}}\mathrm{exp}(2e)}{cC_{\mathrm{h}}+\mathrm{exp}(e)}\theta\label{eq:theta_general}\end{eqnarray}
 where $V$ is the dimensionless applied potential, $R$ is the dimensionless
series resistance, $\Gamma$ is the surface capacity, and $i_{\mathrm{F}}$
is the Faradaic current:

\begin{eqnarray}
i_{\mathrm{F}}(\theta,e) & = & \left(\frac{C_{\mathrm{h}}\exp(0.5e)}{1+C_{\mathrm{h}}\exp(e)}+a\exp(e)\right)(1-\theta)\label{eq:faradic_general}\end{eqnarray}

The parameter values $C_{\mathrm{h}}=1600$, $a=0.3$, $B=6\times10^{-5}$,
$c=1\times10^{-3}$ were optimized \cite{Haim.Lev.ea:92:Journal-Of-Physical-Chemistry:Modeling-periodic-and-chaotic}
to obtain dynamical features similar to experiments. 

In the experiments the oscillators are inherently different because
of electrode heterogeneities \cite{Kiss.Wang.ea:99:Journal-Of-Physical-Chemistry:Experiments-on-arrays-of}
and because of addition of different individual external resistors.
We model the non-identical nature of the oscillators by giving the
oscillators different values for the parameters $R$ and $\Gamma$
in Eqs.~\ref{eq:ni_general} and \ref{eq:theta_general}. (Other
choices are also possible, however, these parameters best approximate
the experiments.\cite{Zhai.Kiss.ea:04:Industrial-Engineering-Chemistry-Research:Emerging-coherence-of-oscillating,Zhai.Kiss.ea:04:Extracting-order-parameters-from-global})
For element $k$ the resistance and the surface capacitance are obtained
using the relationships $R_{k}=(1+\Delta_{k})R_{0}$ and $\Gamma_{k}=(1+\Delta_{k})\Gamma_{0}$,
where $\Delta_{k}$ is a heterogeneity parameter and $R_{0}$ and
$\Gamma_{0}$ are the mean values. We used a fixed value of $\Gamma_{0}=0.01$,
and $R_{0}=20$ throughout this study. We choose a Lorentzian distribution
for $\Delta_{k}$. (For comparison, simulations with global coupling
were also made with a Gaussian distribution.\cite{Zhai.Kiss.ea:04:Extracting-order-parameters-from-global})
The Lorentzian distribution $p(x)=\gamma/\{\pi[(x-x_{0})^{2}+\gamma^{2}]\}$
is characterized by a parameter $\gamma$; $2\gamma$ is the half-width
of the distribution. For a typical value of $\gamma=0.5$, the equation
parameters $R_{k}$ and $\Gamma_{k}$ vary within a range of 5\% of
their means. 

Electrical global coupling of strength $K$ is considered; the model
for the coupled set of $N$ oscillators is then\cite{Zhai.Kiss.ea:04:Industrial-Engineering-Chemistry-Research:Emerging-coherence-of-oscillating,Zhai.Kiss.ea:04:Extracting-order-parameters-from-global}:
\begin{eqnarray}
\frac{de_{k}}{dt} & = & \frac{V-e_{k}}{R_{k}}-i_{\mathrm{F},k}(\theta_{k},e_{k})+\frac{1}{R_{0}}K(e_{\mathrm{mean}}(t)-e_{k})\label{eq:ni_individual}\\
\Gamma_{k}\frac{d\theta_{k}}{dt} & = & \frac{\mathrm{exp}(0.5e_{k})}{1+C_{\mathrm{h}}\mathrm{exp}(e_{k})}(1-\theta_{k})-\frac{BC_{\mathrm{h}}\mathrm{exp}(2e_{k})}{cC_{\mathrm{h}}+\mathrm{exp}(e_{k})}\theta_{k}\label{eq:theta_individual}\end{eqnarray}
 where $R_{0}$ is the mean resistance. Global coupling occurs because
of the presence of mean potential ($e_{mean}(t)=1/N\sum_{k=1}^{N}e_{k}(t)$)
in Eq. \ref{eq:ni_individual}. $K=0$ represents uncoupled oscillators;
$K\rightarrow\infty$ yields maximum global coupling that synchronizes
the oscillators. Global forcing is added through the circuit potential:\begin{equation}
V(t)=V_{0}+b\sin(2\pi f_{\mathrm{F}}t),\label{eq:forcing}\end{equation}
where $V_{0}=15$ is a set potential, $b$ and $f_{\mathrm{F}}=0.0745$
are the forcing amplitude and frequency, respectively. A variable
step size fourth order Runge-Kutta method of Matlab was used to integrate
Eqs.~\ref{eq:ni_individual} and \ref{eq:theta_individual} with
a display step size of $\Delta t=0.5$, absolute error $=10^{-6}$,
relative error $=10^{-3}$. Smaller step sizes and error limits gave
the same results. A transient of $t=10000$ was discarded from each
time series.

The phases and frequencies of simulated oscillators are determined
from time series data of $e_{k}(t)$ with the Hilbert transform method\cite{sync_book}.

\subsection{Results}

\begin{figure}
\includegraphics[%
  clip,
  width=0.9\columnwidth]{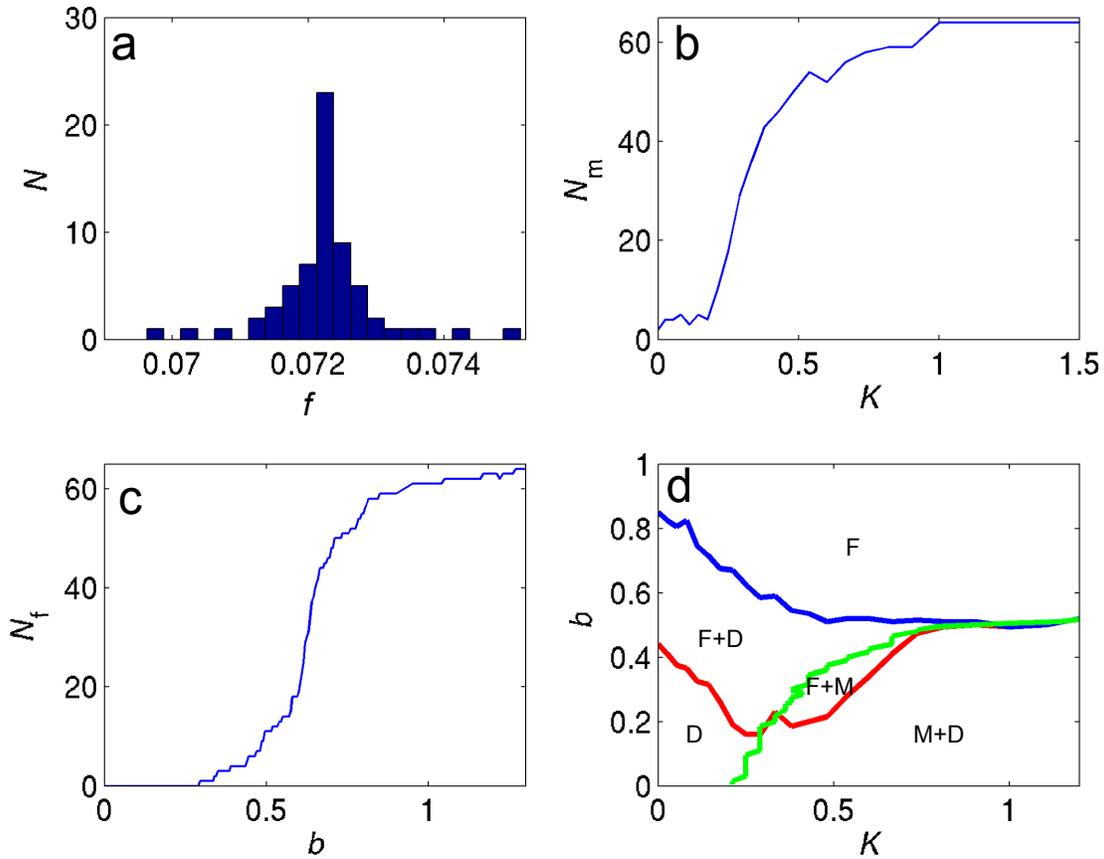}

\caption{\label{fig:num_phase}(Color Online) Numerical Simulations: Mutual
and forced entrainment in a model of sixty-four globally coupled oscillators
with external forcing. a. Natural (inherent) frequency distribution.
b. Mutual entrainment with global coupling only. Number of mutually
entrained oscillators vs. coupling strength $K$. c. Forced entrainment
with external forcing only. Number of oscillators locked to external
forcing signal vs. forcing amplitude $b$. d. Phase diagram showing
forced and mutually entrained regions in the coupling strength $-$
forcing amplitude parameter space. F: Forced entrainment. M: Mutual
entrainment. D: desynchronized state. $f_{\mathrm{F}}=0.0745$.}
\end{figure}

Without global coupling and forcing the oscillators exhibit frequencies
determined by the introduced heterogeneities, as shown in Fig. \ref{fig:num_phase}a.
 Since the heterogeneities are small, the frequencies of the oscillators
follow almost linearly the heterogeneities and thus the frequency
distribution is similar to a Lorentzian distribution. With global
coupling only, a mutually entrained state occurs (see Fig. \ref{fig:num_phase}b)
above a critical coupling strength $K_{c}$\cite{Kiss.Zhai.ea:02:Science:Emerging-coherence-in-population,Zhai.Kiss.ea:04:Industrial-Engineering-Chemistry-Research:Emerging-coherence-of-oscillating,Zhai.Kiss.ea:04:Extracting-order-parameters-from-global,kuramotobook,Winfree:67:J-Theor-Biol:Biological-Rhythms-and-Behavior};
in this mutually entrained state a large fraction of the oscillators
are synchronized. When the coupling strength is increased above $K_{c}$,
the number of mutually entrained oscillators increases until all the
64 oscillators are fully synchronized. With global forcing only, (see
Fig. \ref{fig:num_phase}c), as the forcing amplitude increases more
and more oscillators lock to the forced entrainment state. Note however,
that there is no phase transition, and the increase in the number
of entrained oscillators with forcing amplitude is determined by the
frequency distribution; the large increase at around $b=0.6$ is due
to the large number of oscillators of similar frequencies (peak in
Fig.~\ref{fig:num_phase}a). 

With both global coupling and forcing, in a large fraction of the
parameter space ($K,b$), each oscillator can be classified as belonging
to one of the three major groups: desynchronized (D), mutually entrained
(M), and forced entrained (F) states. Fig.~\ref{fig:num_phase}d
shows the regions in the ($K,b$) parameter space in which these states
can occur. At low forcing and coupling the system is desynchronized
(D). Starting from this desynchronized state with increasing coupling
there is a transition to the occurrence of the mutually entrained
state (M+D in Fig.~\ref{fig:num_phase}b). With increasing the forcing
amplitude in a weakly coupled population, a forced entrainment state
starts to occur (F+D). At very strong coupling, the transition is
similar to the forcing of a single oscillator: the frequency of the
mutually entrained cluster (M) shifts to that of the forcing (F). 

We investigated the dynamics in the middle region of the phase diagram
further, where both forced and mutually entrained states can co-exist.
(Note that this is not hysteresis. In a synchronized state some of
the oscillators are mutually, some are forced entrained; there are
some oscillators that are also entrained but, as we shall see below,
their frequencies are combination of those of the mutualy and forced
entrainment states.) %
\begin{figure}
\includegraphics[%
  clip,
  width=0.9\columnwidth]{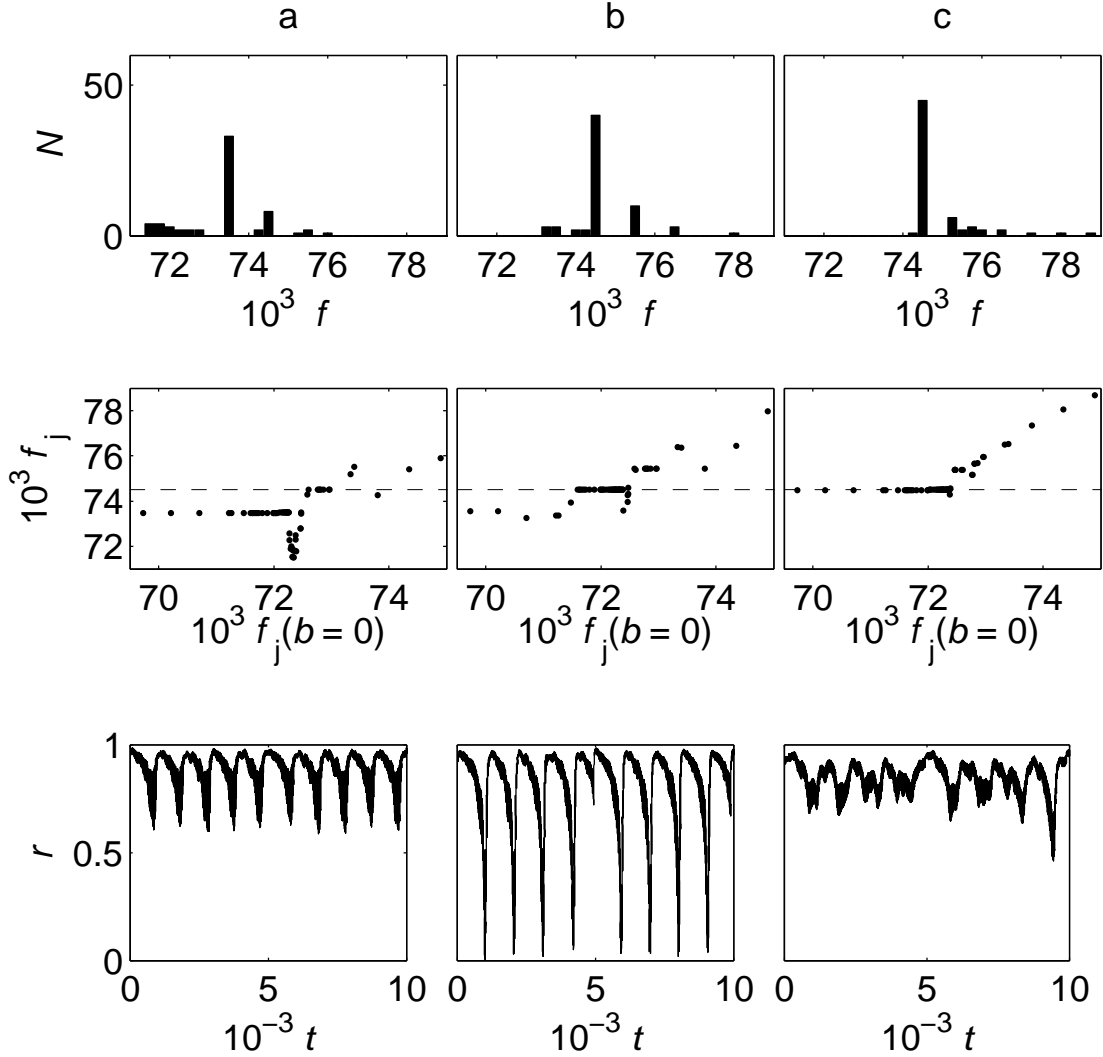}

\caption{\label{fig:num_clustering}Numerical Simulations: frequency distributions
and order parameters at $K=0.67$. Top row: frequency distribution
of oscillators. Middle row: frequency of oscillators vs. inherent
frequency (dashed line denotes the forcedly entrained cluster). Bottom
row: order parameter vs. time. a. Mutual entrainment dominated state
with weak forcing. $b=0.428$. b. Resonance clustering at intermediate
forcing strength. $b=0.443$. c. Entrainment to strong forcing. $b=0.452$. }
\end{figure}
The frequencies and the order of such a state with weak forcing is
shown in Fig.~ \ref{fig:num_clustering}a. Since the forcing is weak,
the large peak in the frequency histogram is associated with the mutually
entrained state, while the smaller peak at the forcing frequency with
the forced entrainment. The frequency vs. inherent frequency graph
(middle) shows that the elements with lower frequencies form the mutually
entrained state; such an asymmetry in the entrainment was also observed
with pure coupling as well. \cite{Zhai.Kiss.ea:04:Industrial-Engineering-Chemistry-Research:Emerging-coherence-of-oscillating}
Elements with frequencies between the mutually and forced entrainment
groups have strongly scattered, lower frequencies; those oscillators
with larger frequencies are not entrained. Because the majority of
the population is mutually entrained, a large value of order parameter
(bottom panel) is observed with small modulation due to forced entrainment
state. 

At a somewhat stronger forcing amplitude, a restructuring of the frequencies
of the oscillators occurs (see Fig.~ \ref{fig:num_clustering}b).
Now the major group is associated with the forced entrainment, however,
the mutually entrained state is difficult to determine since the population
becomes clustered at frequencies with a (dimensionless) spacing of
approximately 0.0009 {[}\textbf{$f=$}\{0.0736 , 0.0745 (forcing),
0.0754 , 0.0764{]}; 8 elements of the population of 64 oscillators
do not belong to these resonant clusters. Resonant clusters seem to
appear as an interaction between the mutual and forced entrainments
and produce a strongly oscillating order parameter. 

As the forcing amplitude is further increased, the forced entrainment
(Fig.~ \ref{fig:num_clustering}b) state dominates in which the elements
around the forcing frequencies along with those of the lower frequencies
are entrained and the elements with higher frequencies are not synchronized.
The order parameter again has a large, close to 1 value, but with
small fluctuations due to the desynchronized elements.

\begin{figure}
\includegraphics[%
  clip,
  width=0.9\columnwidth]{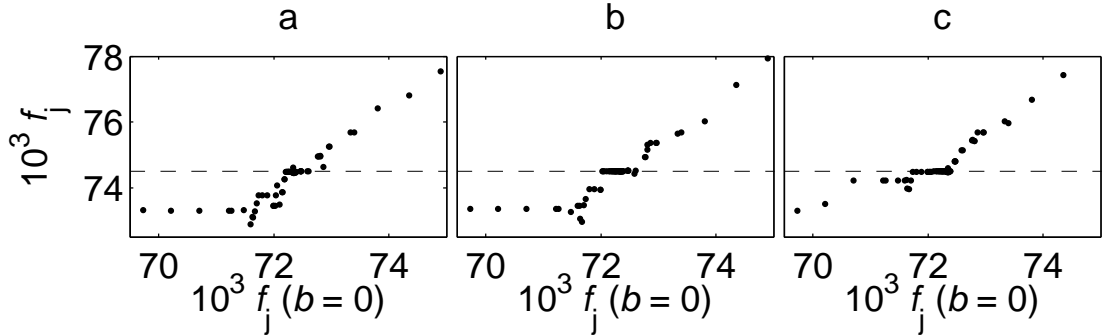}

\caption{\label{fig:num_weak}Numerical simulations: frequencies of oscillators
(with coupling and forcing) vs. inherent frequencies at weaker interactions,
$K=0.54$. a. Weak forcing. $b=0.385$. b. Intermediate forcing strength.
$b=0.398$. c. Strong forcing. $b=0.410$. }
\end{figure}

Resonance clustering to a lesser extent also appears at weaker coupling
strength as it is shown at three forcing amplitudes in Fig.~ \ref{fig:num_weak}.
 At these conditions (Figs.~ \ref{fig:num_weak}ab), the elements
do form various frequency clusters, however, there may be quite a
few clusters with smaller number of elements and the spacing between
the clusters is small. These states also exhibit oscillating order
(not shown in the figures). At a state very close the destruction
of mutual entrainment (Fig.~ \ref{fig:num_weak}c) the frequencies
of mutual and forced entrainment states are very close and frequency
clustering appears as discretized frequencies of groups of 1-3 oscillators
only.

\begin{figure}
\includegraphics[%
  clip,
  width=0.9\columnwidth]{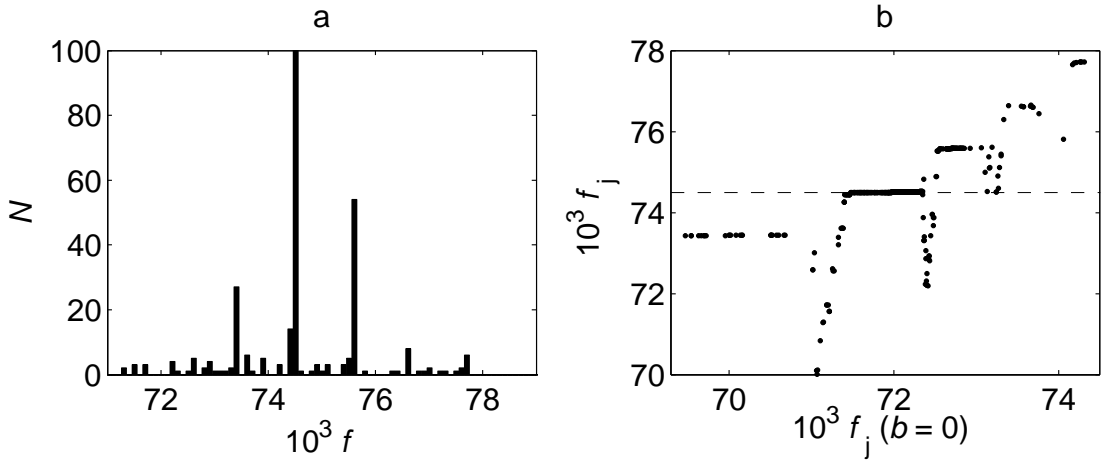}

\caption{\label{fig:num_512}Numerical simulations: population of 512 oscillators
exhibiting resonance clustering at $K=0.67$ and $b=$0.47. a. Frequency
distribution of oscillators. (For $f=0.075$ there is a large
peak with $N=325$, therefore, a zoom is shown) b. Frequency of oscillators
vs. inherent frequency (dashed line denotes the forcing frequency).}
\end{figure}
 Fig. \ref{fig:num_512} shows the simulation results with a population
of 512 oscillators at inherent frequency distribution, coupling strength,
and forcing amplitude similar to those investigated with 64 oscillators.
We observed five frequency clusters with approximate spacing of 0.001
{[}at frequencies $f$=\{0.0734, 0.0745 (forcing), 0.0756, 0.0766,
0.0777\}{]}. This 5-cluster state with $N=512$ is reminiscent of
the 4-cluster state with $N=64$ (Fig. \ref{fig:num_clustering}b),
however, one additional cluster could be resolved at $f=0.0777$.
Thus, the fine structure of resonance clusters can be better seen
with larger populations.

\section{Discussion}

Clustering is a behavior of a population of oscillators through which
dynamically differentiated elements form. The classification of clusters
\cite{Mikhailov.Showalter:06:Phys-Rep:Control-of-waves-patterns}is
based on whether it affects mainly the phases (phase clusters)\cite{Golomb.Hansel.ea:92:Phys-Rev:Clustering-in-globally-coupled,Okuda:93:Physica-D:Variety-and-generality-of}
or the amplitudes (amplitude clusters)\cite{Hakim.Rappel:92:Phys-Rev:Dynamics-of-globally-coupled}
of the periodic oscillators. Chaotic systems are particularly inclined
for producing clusters at coupling strengths weaker than those required
for identical synchronization\cite{Kaneko:89:Physical-Review-Letters:Chaotic-but-regular-posi-nega,Kaneko.Tsuda:01:Complex-systems,Wang.Kiss.ea:00:Chaos:Experiments-on-arrays-of}.
In this study, we observed frequency clusters in a forced, weakly
coupled population of limit-cycle electrochemical oscillators with
unimodal heterogeneities. In addition to the trivial mutually and
entrained clusters, groups of elements are obtained at other, discrete
frequency values. We also showed that at coupling strengths and forcing
amplitudes with well-defined resonance clusters the frequencies are
equally spaced and follow the relationship found by Sakaguchi \cite{Sakaguchi:88:Progress-Of-Theoretical-Physics:Cooperative-phenomena-in-coupled}:
$\omega_{n}\cong n\omega_{1}-(n-1)\omega_{\mathrm{F}}$. (Such spectrum
was analyzed by linear stability of the forced entrainment state.\cite{ott_sakaguchi})
With weaker coupling strengths a large number of clusters with small
number of elements were observed whose frequency was not greatly modified
from the natural frequencies. We note that frequency discretization
through a different mechanism, \emph{viz}., through large delays in
the coupling term, was also observed in a pair of lasers \cite{Wunsche.Bauer.ea:05:Phys-Rev-Lett:Synchronization-of-delay-coupled-oscillators}and
in a pair chaotic R\"ossler oscillators\cite{Yanchuk:05:Phys-Rev-E:Discretization-of-frequencies-in}. 

Along with the occurrence of frequency clusters, a strongly oscillating
order parameter was observed; the order decreased to very low values
and thus intermittent loss of the overall rhythm could be observed.
The loss of the overall rhythm may have implications in the dynamics
of biological systems where the rhythm could be either essential or
pathological \cite{winfreebook}. 

A biological example in which coupling and forcing play a role is
the circadian master clock in the brain\cite{Reppert.Weaver:02:Nature:Coordination-of-circadian-timing};
the suprachiasmatic nuclei consist of heterogeneous, circadian oscillators
that are entrained by cell interactions and by external light. The
SCN is strongly heterogeneous and the interaction of the core, being
entrained by light, and the shell (composed of mostly self-oscillating
neurons) is a complex issue\cite{Antle.Silver:05:Trends-in-Neurosciences:Orchestrating-time-arrangements-of,Nakamura.Yamazaki.ea:05:J-Neurosci:Differential-response-of-Period};
resonance clustering could play a role in the dynamical behavior of
the different regions. Resonant interactions between oscillators have
been considered as well in the context of information processing with
oscillatory units \cite{Hoppensteadt.Izhikevich:98:Biosystems:Thalamo-cortical-interactions-modeled-by,Hoppensteadt.Izhikevich:99:Phys-Rev-Lett:Oscillatory-Neurocomputers-with-Dynamic};
the cortex is considered as a weakly coupled oscillator population
forced by the thalamic input. 

As far as the importance of resonance clustering in biological networks
of rhythmic elements is concerned, we shall consider that we investigated
a global coupling topology. Although such global, all-to-all coupling
is not very likely in biological systems, sometimes a network of oscillators
can be approximated by global coupling. A fundamental communication
mechanism of bacteria, quorum sensing, is often modeled by global
interactions; such a mechanism was shown to be able to produce synchrony\cite{Garcia-Ojalvo.Elowitz.ea:04:Proc-Natl-Acad-Sci:Modeling-synthetic-multicellular-clock}.
Synchronization of circadian cells was modeled by a global interaction
mechanism based on an argument that the spatial transmission of neurotransmitters
released by each cell is fast compared to the timescale of oscillations\cite{Gonze.Bernard.ea:05:Biophys-J:Spontaneous-synchronization-of-coupled}.
Neural networks with electrically spiking neurons are also often can
be considered as a population of weakly, coupled globally coupled
oscillators\cite{Kori:03:Phys-Rev-E:Slow-switching-in-population}.
Even when network coupling occurs, the depth of the network is often
not very large when external forcing is effective\cite{Kori.Mikhailov:04:Phys-Rev-Lett:Entrainment-of-randomly-coupled}
thus global coupling approximation and resonance clustering can play
a role in generation of collective dynamics with coupling and forcing.

Some features of the cluster interactions such as generation of the
resonant clusters could also be observed in a population with bimodal
heterogeneities; the forced system can be regarded as a simplified
case of bimodal system where one special mode has standard deviation
of zero and its strength can be directly controlled. Frequency clustering
and generation of complex collective signal were found with the global
coupling of periodic electrochemical oscillator populations with bimodal
natural frequency distributions\cite{Mikhailov.Zanette.ea:04:Proceedings-of-National-Academy:Cooperative-action-of-coherent};
merging and splitting of clusters occurred on the way to the final
synchronized state with increasing the coupling strength. Such bimodal
(and multimodal) populations may occur in biological systems that
are composed of broadly heterogeneous cell groups. 

Although we investigated resonance clustering here with electrochemical
oscillators, similar dynamical differentiation mechanisms are expected
to occur in a variety of rhythmic multicellular systems under the
co-operative effects of coupling and forcing; the synchronized groups
of elements would contribute to the formation of multi-structured
hierarchical organizations often seen in complex systems\cite{emergencebook,kaneko_book_system}.

\begin{acknowledgments}
Financial support from the National Science Foundation \textbf{(}CBET-0730597)
is acknowledged. 
\end{acknowledgments}

\end{document}